\documentclass{article}

\usepackage{PRIMEarxiv}

\usepackage[utf8]{inputenc} 
\usepackage[T1]{fontenc}    
\usepackage{hyperref}       
\usepackage{url}            
\usepackage{booktabs}       
\usepackage{amsfonts}       
\usepackage{nicefrac}       

\usepackage{amsmath}
\usepackage{microtype}      
\usepackage{lipsum}
\usepackage{fancyhdr}       
\usepackage{graphicx}       
\graphicspath{{media/}}     

\title{Validation of Thin-foil proton recoil neutron spectrometer prototype for application in high yield DT fusion devices}

\author{
  B. Marcinkevicius, E. Andersson Sund\'en, G. Ericsson, A. Hjalmarsson \\
  Affiliation \\
  Uppsala University \\
  Uppsala, Sweden\\
  \texttt{bmarcinkevicius@gmail.com} 
}

\begin{document}
\maketitle

\begin{abstract}
The use of Thin-foil Proton Recoil (TPR) spectrometers for application in neutron spectroscopy is of high relevance for future fusion devices such as ITER, where neutron spectroscopy will play a crucial role in fuel content monitoring. Existing research based on simulations of the performance of TPR spectrometers at ITER has demonstrated positive results. However, experimental validation of the simulations would greatly benefit the reliability of conclusions. In this study, we designed and constructed a prototype TPR neutron spectrometer and employed a DT neutron generator as a neutron source to perform measurements. We compared the experimental results with the simulation results using the  Geant4 model of the experiment. The simulation and experimental results match within  silicon detector intrinsic energy resolution. This approach ensures the experimental validation of the Geant4 based simulations of the TPR spectrometer.
The experimental results demonstrated the feasibility of utilizing nuclear reactions measured in silicon detectors, specifically $^{28}$Si(n,d) and $^{28}$Si(n,$\alpha$), for energy calibration purposes. A comparison of the experiment and  the simulation shows that the mean peak energy and full width at half maximum are within 150 keV. The calculated detector efficiency underestimates the experimentally determined efficiency up to 33\%. Discrepancies in the measured energy spectrum indicate the need for a more refined model and experiment control.
Overall, the successful validation of the developed Geant4 simulation model against the experimentally measured energy spectra increases confidence in the applicability of such simulation results in other devices. The demonstrated energy calibration highlights new possibilities for neutron spectrometer monitoring during operation at ITER. 

\end{abstract}


\section{Introduction}
\label{sec:intro}
A fusion reactor ITER~\cite{iter} is under construction in France to demonstrate the possibilities of fusion as a carbon-free source of energy. The reactor will require extensive diagnostic systems for control and research purposes. One of the key requirements is to monitor the fuel ion temperature $T_i$ and the fuel ion ratio $n_t/n_{tot}$ during operation to maintain the optimal fuel burning conditions. This could be provided by a neutron spectrometer system as demonstrated at JET \cite{sjstrand_new_2006,sunden2009,Hjalmarsson2003,gatu_johnson_25-mev_2008}.  For these purposes, a neutron spectroscopy system is under development~\cite{scholz_conceptual_2019}, and an important part  is the Thin-foil Proton Recoil (TPR) neutron spectrometer. The capabilities and limitations of the TPR spectrometer for application in fusion devices has been investigated by simulations~\cite{scholz_conceptual_2019,Marcinkevicius_2019,Marcinkevicius2018,Marcinkevicius2019}. Experimental validation of these simulations would increase the confidence in the performance of the TPR spectrometer at ITER.

The purpose of this paper is to examine whether simulations based on the Geant4 toolkit~\cite{geant_2ref,ALLISON2016}  can be validated by an experiment. Consequently, a prototype of the TPR spectrometer was designed and tested using a DT neutron generator with a neutron peak energy of 14~MeV similar as anticipated at ITER. 

The details of a prototype TPR spectrometer design are given in Section~\ref{sec:tpr_system}, which contains a description of the working principles of the TPR spectrometer, the implemented TPR spectrometer design, information about the data acquisition system, description of the energy calibration experiment, and results of the energy calibration.
The neutron experiment is given in Section~\ref{sec:exp}, which describes the details of the neutron generator, experimental conditions, data analysis principles, and results of the measured coincidence and anti-coincidence energy spectra. 
The simulations performed are given in Section~\ref{sec:simulations} describing the implementation of the experimental geometry in the Geant4 model and results covering: energy coincidence spectra, region-of-interest estimation, energy anti-coincidence spectra and energy calibration correction. A comparison of the experiment and simulation results is presented in Section~\ref{sec:results}  and a summary of findings and future prospects is discussed in Section~\ref{sec:conclude}.

\section{Thin-foil proton recoil neutron spectrometer prototype}
\label{sec:tpr_system}
\subsection{TPR neutron spectrometer}
\label{subsec:tpr}

\begin{figure}[!ht]\centering
	\includegraphics[width=0.67\linewidth]{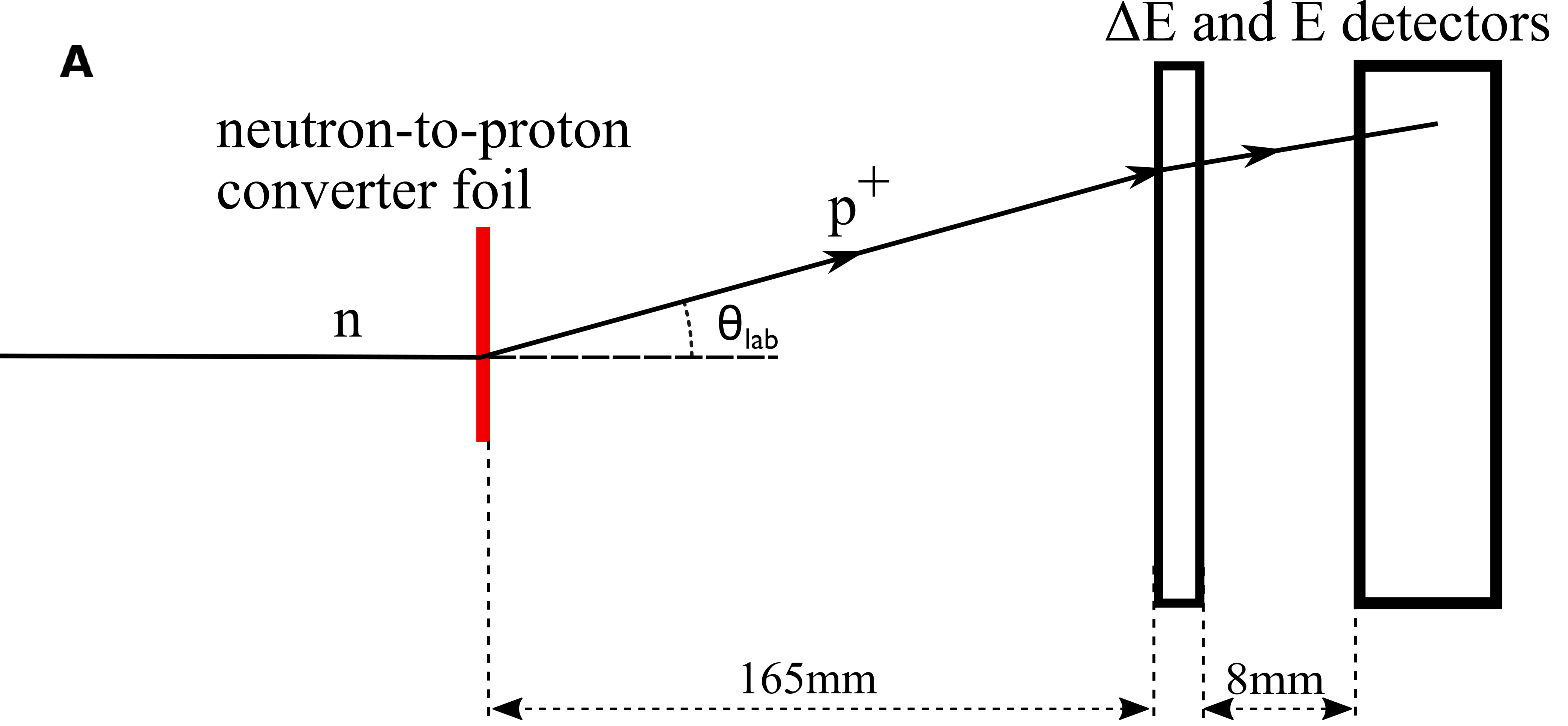}
	\includegraphics[width=0.28\linewidth]{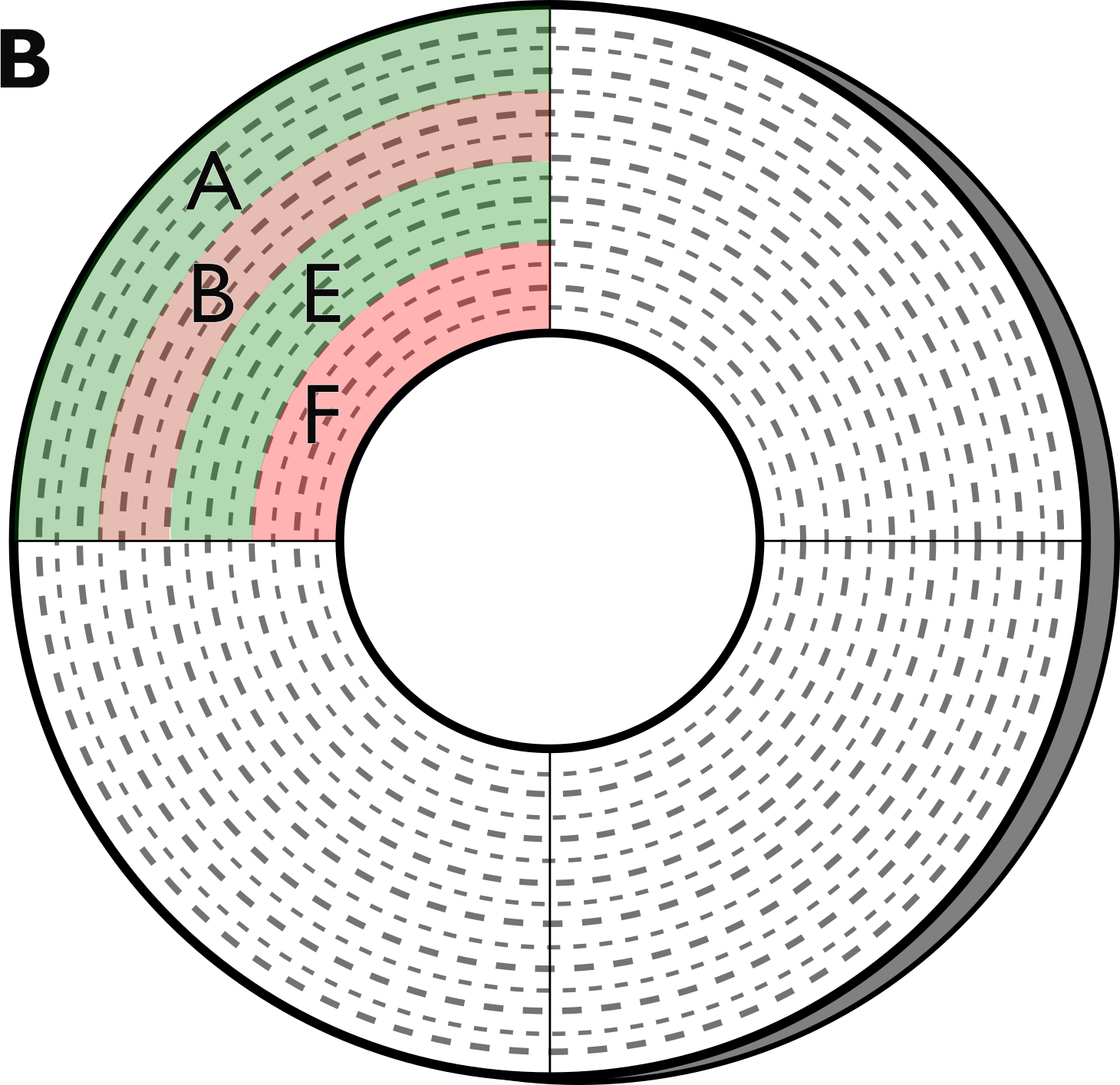}
	\caption{A - Sketch of prototype TPR spectrometer showing operating principles and relevant dimensions.  The figure is not to scale. B - sketch of the $\rm \Delta E$ segmented silicon detector; shaded areas in green an red (A, B and E,F) correspond to the detector coupling to the acquisition system's channels A, B, E and F. The E detector is coupled to acquisition channels C, D and G, H in an identical layout to the $\rm \Delta E$ detector.}
	\label{fig:tpr}
\end{figure}

The thin-foil proton recoil spectrometer relies on the basic principles of neutron elastic scattering on hydrogen (protons) in a neutron-to-proton converter foil. In such reactions , the incoming neutron can knock out a proton which in turn can deposit energy in suitably positioned detectors (see figure~\ref{fig:tpr} panel A). The proton energy can be related to the neutron energy through the scattering angle: $E_p = E_n cos^2(\theta_{lab})$ where $ E_p$ is the recoil proton energy, $ E_n$ is the incoming neutron energy, and $ \theta _{lab}$ is the angle of scattering between neutron and proton in the laboratory system. 

\begin{figure}[!ht]\centering
	\includegraphics[width=0.45\textwidth]{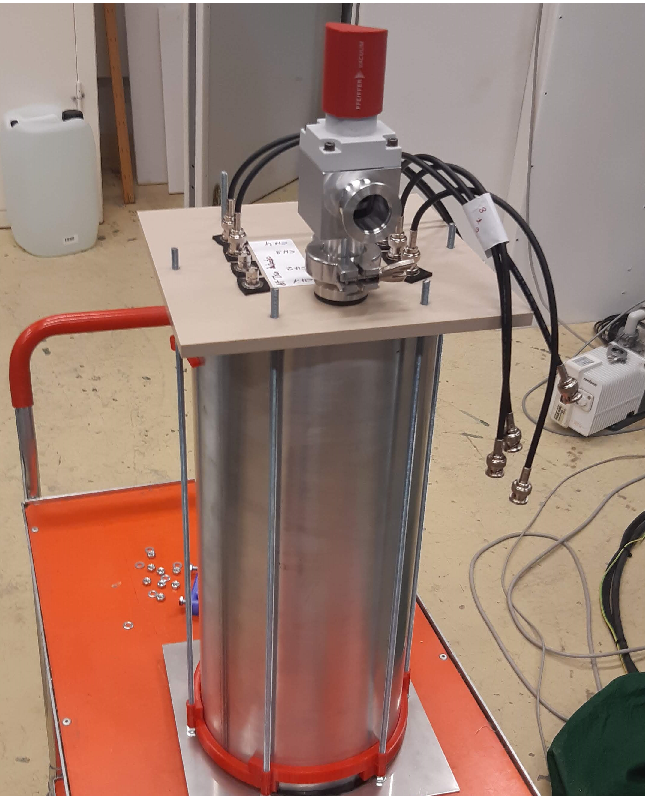}
	\caption{Photo of vacuum chamber used for the tests placed vertically. The front steel plate is visible at the bottom, the light-tight plastic plate is at the top eight BNC connector feed-thrus.  The cylindrical casing is made out of aluminum. }
	\label{fig:vacuum_chamber}
\end{figure}

The prototype TPR spectrometer used in this study consists of two segmented, annular silicon detectors  ($\rm \Delta E $ and E) of 0.3 and 1~mm thickness, respectively, placed concentrically in a proton telescope configuration. The distance (surface to surface) between the neutron-to-proton converter foil and the $ \Delta $E detector is 165~mm while the distance between the two detectors is 8~mm as shown in figure~\ref{fig:tpr}.  The converter foil is made of a low-density (0.92~$ \rm g / cm ^{-3}$) polyethylene has a radius of 21~mm  and thickness of  0.239~$ \pm $~0.004~mm. Each detector has 16 radial segments and is equidistantly divided into 4 quarters; the inner and outer radius of the annulus are 24 and 48~mm. The active volume of the detector has inner and outer radius of 32 and 48~mm respectively. Each segment has a junction area and the separation between the radial segments of 0.1~mm has been reported~\cite{torresi_influence_2013} for similar detectors as used in this experiment~\cite{micron}. It can be expected that detector sensitivity will be reduced in this are and as a result this will influence the total collected charge and affect the pulse height spectra. 

The spectrometer was positioned inside an aluminium cylindrical vacuum chamber. The front plate of the chamber was constructed from steel with dimensions of 3.4 mm x 250 mm x 250 mm, while the back plate was made from light-tight plastic, measuring 3 mm x 250 mm x 250 mm. The internal pressure of the vacuum chamber was maintained at 0.01~mbar to maintain the loss of proton energy due to residual air at acceptable levels (< 1~keV). Steel was selected as a material for the front panel instead of aluminium in order to avoid the low threshold (around 3~MeV) nuclear reaction $ \rm ^{27}Al(n,p)$ which might otherwise obscure the signal produced by recoil protons.

Protons of energies above 6.1~MeV  impinging perpendicular to the detector surface can penetrate the $\rm \Delta E$ detector and can deposit energy in both $\rm \Delta E$ and E detectors. Protons of energy above 14.1~MeV would punch-through both of the detectors and would deposit only part of their energy. Consequently, the spectrometer operational range is limited to proton energies of 6.1 - 14.1~MeV which corresponds to a similar range in neutron energies, depending on the scattering angle and trigger threshold. 

The prototype design was selected to achieve a sufficient number of events given the experimental constrains. Energy resolution was optimised using the same methods as in previous studies~\cite{Marcinkevicius2019} using the existing constraints imposed by the experimental setup. The simulation was performed assuming a parallel monoenergetic neutron beam, ideal detectors and considering only elastic scattering on hydrogen in the foil. The estimated spectrometer energy resolution is  $\frac{\delta E}{E} = 8.2\%$, Full Width at Half Maximum (FWHM), at a neutron energy of 14 MeV, based solely on the spectrometer's geometry.


\subsection{Acquisition system}
\label{subsec:acc_system}
An acquisition system has been designed for the TPR spectrometer. 
Both the $\rm \Delta E$ and E detectors were connected to a CAEN1422~\cite{CAEN} charge sensitive pre-amplifier. Pre-amplifier signals were collected using two ADQ412 12-bit 4 channel digitisers~\cite{SPDevices} and stored for later off-line data analysis to determine the pulse height amplitude and pulse arrival time. The digitizers were nearly identical with only difference that digitizer ADQ412-0901 was lacking the offset voltage capability, which was available for digitizer ADQ412-3920 . 

\begin{figure}[!ht]\centering
	\includegraphics[width=0.99\textwidth]{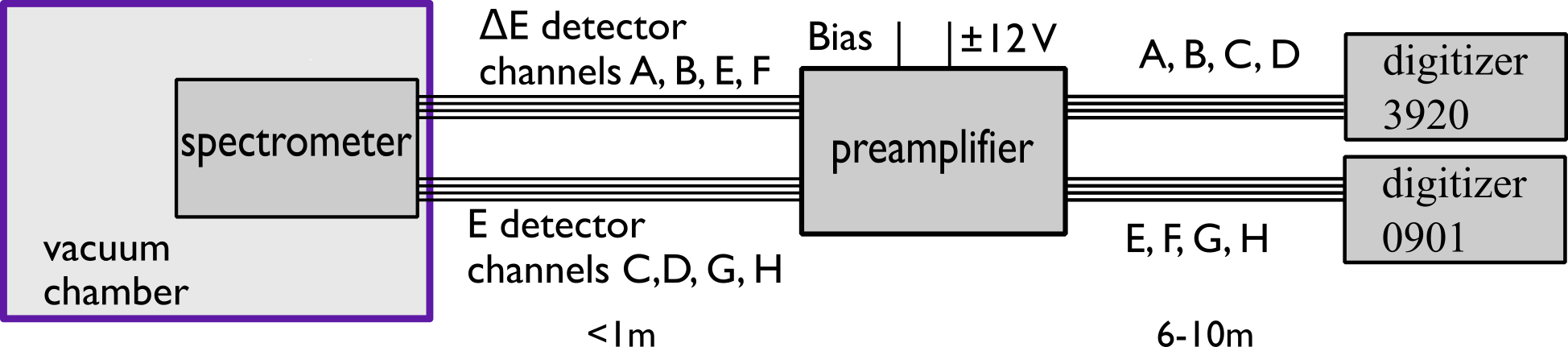}
	\caption{Spectrometer acquisition system scheme; preamplifier - CAEN1422;  digitizer - ADQ412.  The approximate distance between the parts of the acquisition system are marked at the bottom of the figure.  }
	\label{fig:electronic_sketch}
\end{figure}

Only one quarter of each detector was used and 4 radial segments were combined and coupled to a common acquisition channel (see figure~\ref{fig:tpr} right panel). This ensured that the capacitance of the connected detector was within the preamplifier specifications. Each detector was connected to four channels of CAEN1422\cite{CAEN} charge sensitive preamplifier (see figure~\ref{fig:electronic_sketch}). The detector  $ \rm \Delta E$  is connected to channels A, B and E,F and detector E is connected to channels C, D and G, H.  Channels A and C correspond to the outermost detector segments. Channels A,B,C and D were connected to ADQ412-3920 while channels E,F,G and H were connected to an ADQ412-0901 digititzer.  Because ADQ412-0901 lacked the voltage offset capabilities the a 3dB attenuators were used for channels G,H to stay within the digitizer input voltage range and avoid waveform clipping. The cables used to connect the detectors to the preamplifiers (approximately 1~m) or preamplifier to digitizer (6-10~m) were consistent length. Precise length of the cable was not fundamental to the measurement.  The bias voltage for the $ \rm \Delta E$ and E detectors were 60~and~90~V, respectively, to fully deplete the silicon detectors. 

During the energy calibration measurements, a threshold trigger was active on each of ADQ412 accquisition channel individually and each event would be recorded independently. 
During the neutron experiment, two trigger modes were used. Mode one: threshold trigger active on all channels, and once a triggered event is detected in any channel, all channels are read out and recorded. Mode two: threshold trigger active on channels connected to the E detector; after each triggered event, all channels are recorded. Mode two has a lower data transfer rate because protons that stop in the $\rm \Delta E$ detector do not generate a trigger event and thus fewer events are recorded compared to Mode one. During data transfer to the storage device, the digitiser would be idle. Consequently, the second trigger mode is favourable and avoids the bottleneck introduced by the data transfer to the storage device. 
 The used electronics did not have a specific timing shaping circuit to provide accurate trigger times for each of the channels. Instead, the digitiser trigger marked the start of an event.  The system operates consistently across all channels at a time step of 0.5~ns, equivalent to a digitiser sampling rate of 2~GHz for the purposes of timing accuracy. The waveform sampling rate was 15625~kHz, which is sufficient due to long waveform duration of up to ~200~$ \rm \mu seconds$ .
\subsection{Energy calibration}
\label{sec:energy_calibration} 

The energy calibration was carried out in the Tandem laboratory~\cite{tandem} at Uppsala University using a proton beam of multiple energies.  The experimental set-up and data reduction was nearly identical to the previous calibration~\cite{Marcinkevicius2018} and a brief review with relevant differences are given below.


\subsection{Data reduction and Calibration Results}
\begin{figure}[!ht]\centering
    \centering
    \includegraphics[width=0.95\textwidth]{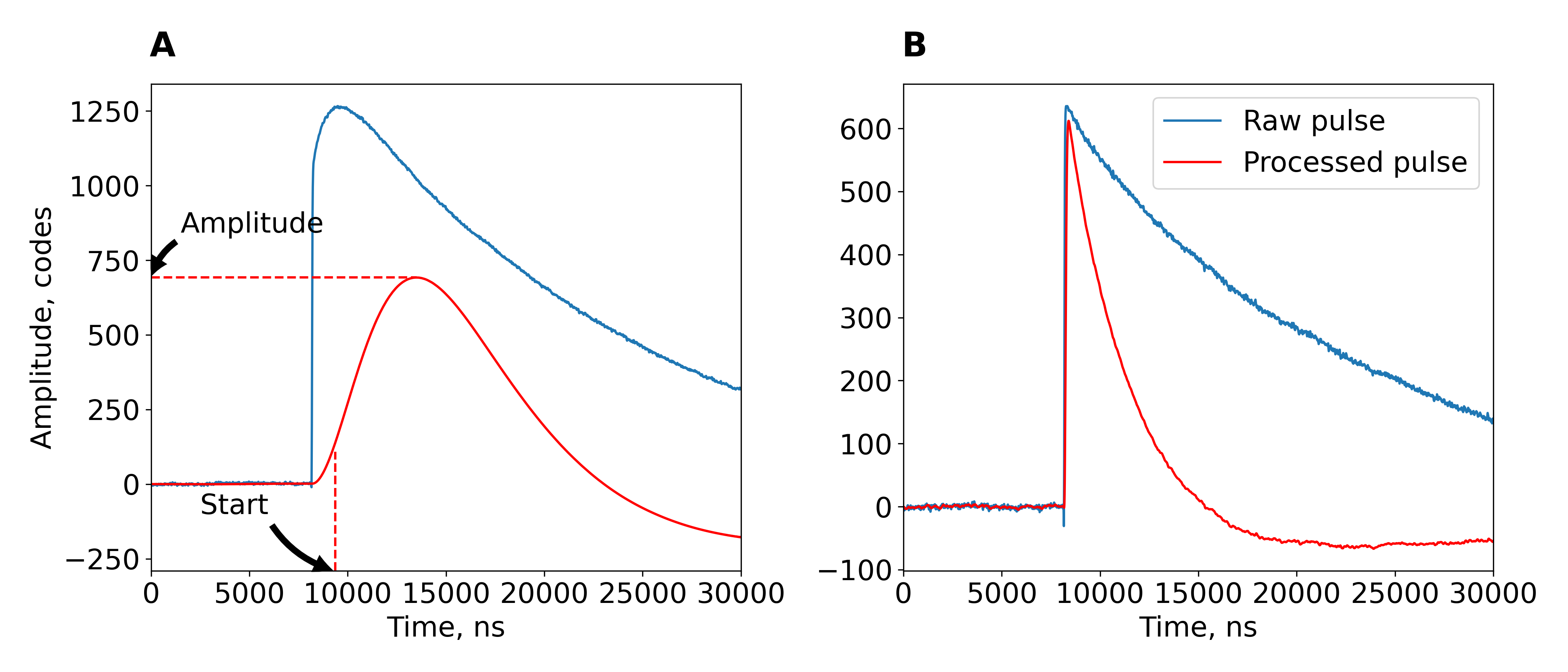}
    \caption{ A - Example of raw pulse and processed pulse after applying filter and interpolating for the E detector. Markings show the amplitude and the start time of the pulse; B - Example of raw pulse and pulse after applying filter and interpolating for the $\rm \Delta $ E detector.}
    \label{fig:signal_example}
\end{figure}

Each recorded event was analysed to extract the amplitude in units of digitiser codes and then construct pulse height spectra. First, pulses were corrected for baseline shifts using the mean amplitude  before the arrival of each pulse. Second, a 2-nd order bandpass filter and Whittaker-Shannon~\cite{Shannon1949} interpolation were applied to each pulse.  For the E detector the bandpass filter had lower cutoff frequency at 5 kHz and upper frequency cutoff of 20 kHz. For the $\rm \Delta E$ detector, the cutoff frequencies were 10 kHz and 1000 kHz.  This minimised the influence of random fluctuation on the pulse amplitude and allowed for a more accurate estimation of pulse height amplitude. An example of a pulse acquired in the detectors after baseline correction is shown in figure~\ref{fig:signal_example} where the maximum amplitude was extracted from the interpolated pulse. The time at 20~\% of maximum amplitude was used as the pulse start time, as shown in figure~\ref{fig:signal_example}A. The $\rm \Delta E$ detector has significantly better timing characteristics compared to E detector. 

\begin{figure}[!ht]\centering
	\includegraphics[width=0.99\textwidth]{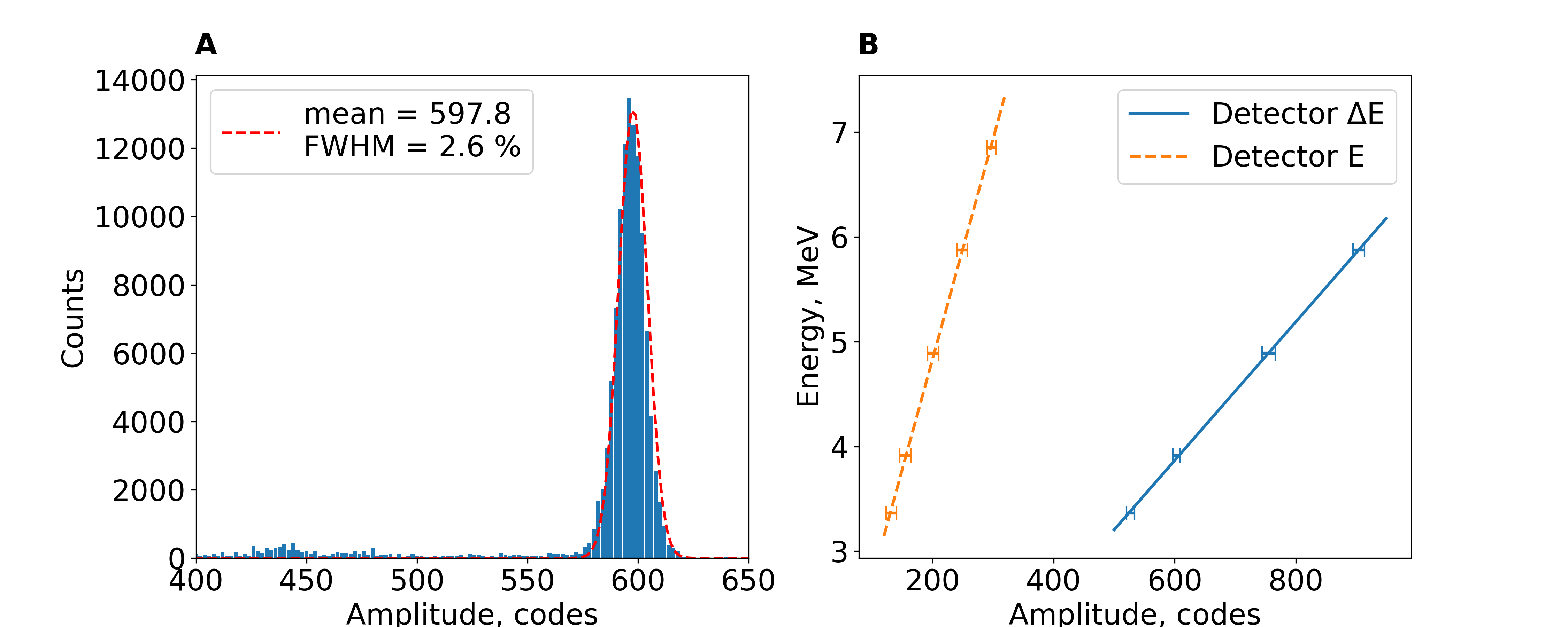}
	\caption{The figure shows an example of the energy calibration steps. A - example of calculated PHS (blue bars)  for 4~MeV proton beam energy as well as fit (red dashed line) for energy calibration purposes;  $\rm \Delta E$ detector. B -  example of energy calibration curve for one channel for each of the detectors. Each marker corresponds to mean and $ \sigma$ derived from the fit to the PHS as in panel A.}
	\label{fig:erg_calib}
\end{figure}

 Finally, a histogram of all pulse amplitudes was constructed to form a pulse height spectrum (PHS); an example of a PHS from the $\rm \Delta E$ detector energy calibration measurement is shown in figure~\ref{fig:erg_calib}A. The mean amplitude and the FWHM of the PHS corresponds to the impinging proton beam energy and detector resolution, respectively. For each of the acquisition channels and proton energies, the mean and FWHM were estimated using a Gaussian fit to the data. A software threshold was used and pulses below the amplitude of 50 codes were excluded from the PHS analysis, this approximately corresponds to 830~keV  and 200~keV for detectors E and $\rm \Delta E$ respectively.

The determined mean amplitude together with the FWHM was used to relate the proton energy deposition to the amplitude using a linear fit. A linear fit has been performed for each of the digitizer channels, and an example of such a fit is shown in figure~\ref{fig:erg_calib}B. The energy calibration is highly linear for all channels, as expected for silicon detectors. The $ \rm \Delta E$  detector performed better compared to the E detector for all proton energies in terms of energy resolution. The $ \rm \Delta E$  detector FWHM was better than 220keV for all experimental points, for the E detector the FWHM varied between 250 and 700 keV depending on the proton energy and acquisition channel. The leakage current was significantly lower for the $ \rm \Delta E$  detector as well. Higher leakage current is indication of higher noise in those detector channels~\cite{Lutz2007}.

Due to the low count rate  (<200~Hz) during the energy calibration the pile-up probability was negligible and was not taken into account during the data analysis. 


\newpage
\section{Neutron Experiment}
\label{sec:exp}
The neutron experiment was carried out in the Laboratory of Applied Nuclear Physics of Lund University. The section further provides the general information about the neutron generator, experimental setup, data reduction,  and main findings of the experiments

\subsection{Neutron Generator}

In this experiment, we used a DT neutron generator manufactured by Sodern~\cite{sodernNeutronTubes} of type Genie-16. The generator accelerates a deuterium beam up to 120~keV which then hits a target enriched with tritium. This, in turn, produces neutrons of approximately 14~MeV via the dominating DT fusion reactions (equation~\ref{eq:dt}) as in fusion reactors.

\begin{equation}
\label{eq:dt}
  \rm  ^2 _1D + _1 ^3T \xrightarrow{} _2 ^4He + n + 14~MeV  
\end{equation}

The neutron can originate from less probable reactions such as TT and DD; furthermore, because of reaction kinematics, the emitted neutron energy will have an angular dependence.  For the purposes of this experiment, the secondary reactions, TT and DD, have negligible influence. The estimation of the angular dependence is discussed further in section~\ref{sec:simulations}. The generator was absolutely calibrated at the factory and provides a neutron yield of $\rm 4.7 \times 10^8 \frac{neutrons}{s}$ in continuous operation mode. 

\subsection{Experiment}
\begin{figure}[!ht]\centering

	\includegraphics[width=0.5\textwidth]{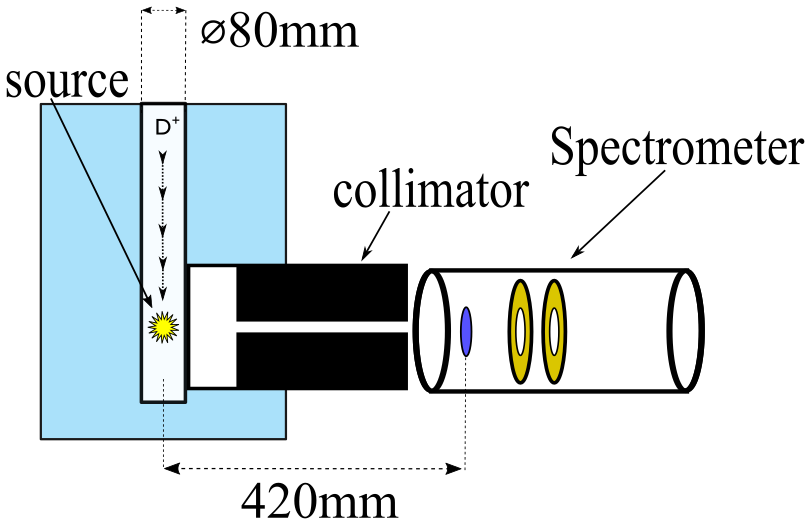}

	\caption{Sketch of the neutron experiment (not to scale). Neutron-to-proton conversion foil is marked dark blue, Si detectors are marked in gold colour, collimator in black. The source is surrounded with water (light blue) acting as a neutron shielding.  $ \rm D^+$ marks the deuterium beam. }
	\label{fig:experiment_setup}
\end{figure}

\begin{figure}[!ht]\centering

	\includegraphics[width=0.98\textwidth]{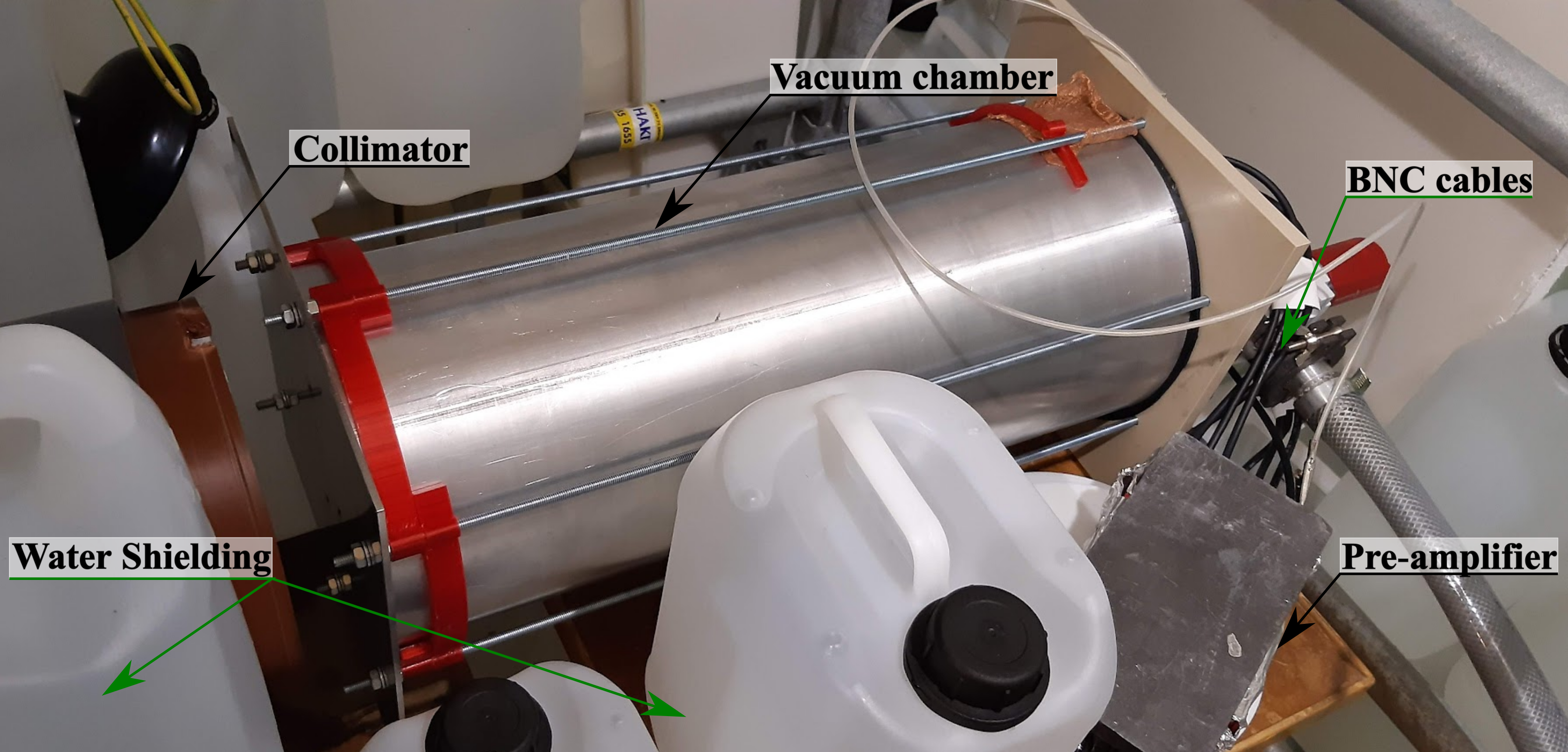}

	\caption{Photo of the vacuum chamber with neutron spectrometer in the place of the experiment. The main parts visible are labelled in the figure.  }
	\label{fig:experiment_photo}
\end{figure}

A schematic drawing of the experiment is shown in figure~\ref{fig:experiment_setup} with the neutron generator ("source") placed inside a cylinder pipe 80~mm in diameter and surrounded by water shielding. $ \rm D^+$ marks the deuterium beam impinging on the target producing neutrons. The generator shielding has a circular opening of 191~mm diameter which was used to hold a paraffin collimator. The collimator is 320~mm in length with inner and outer radius of 18~mm and 95~mm respectively. The spectrometer was placed as close to the far end  of the collimator as physically possible. The spectrometer line-of-sight is perpendicular to the deuterium beam direction. The distance from the neutron source to the n-p conversion foil is approximately 420~mm.  There is a small gap between the collimator and the neutron generator target due to experiment limitations. In addition, for shielding purposes the  neutron generator is covered in borated plastic, and the experimental hall is surrounded by concrete walls, borated plastic and additional water shielding which is not depicted in the figure. A photo of the spectrometer vacuum chamber and spectrometer in measurement position  is presented in figure~\ref{fig:experiment_photo}. 

Background measurements were performed while the neutron generator was not operating. The spectrometers background count rate was orders of magnitude lower than  during the experiment and its influence on the experimental results is negligible. 

\subsection{Data analysis}
The baseline correction, peak amplitude extraction was performed in the same way as for the energy calibration experiment (see section \ref{sec:energy_calibration}). The coincidence window was 7.5~$\rm \mu s $, due to poor timing performance of the E detector. The calculated pulse heights were converted to energy by using the energy calibration curves as shown in figure \ref{fig:erg_calib}. The maximum count rate during the experiment was below 300 Hz thus random coincidence event contribution to the final energy spectra should be negligible. 


\subsection{Results}
\label{sub:exp_results}
\subsubsection{Coincidence Measurement}
\begin{figure}[!ht]
    \centering
    \includegraphics[width=0.99\textwidth]{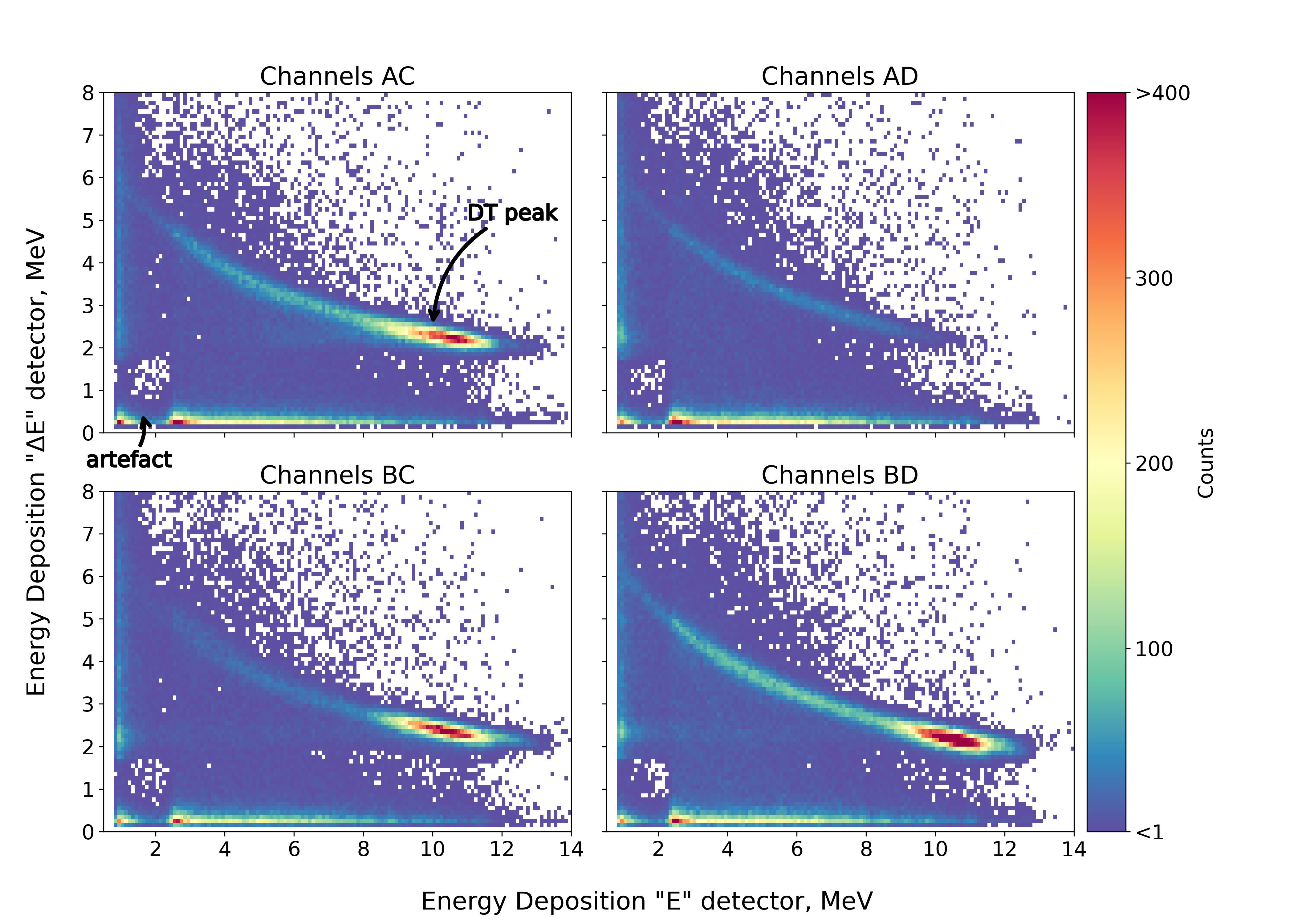}
    \caption{Coincidence energy deposition in the spectrometer for different channel. Channel combinations AC, BC, BD contain a prominent peak associated with the 14~MeV neutrons, marked "DT peak". Channel combination AD does not contain this feature. Artefact marks energy cut due to acquisition trigger level. The energy bin width is 100~keV.  }
    \label{fig:coincidence_erg_dep}
\end{figure}

The energy deposition in the $ \rm \Delta E$ and $ \rm E$ detectors was combined to create the coincidence $ \rm \Delta E-E$  energy deposition spectra  for each of the acquisition channel combinations and is shown in figure~\ref{fig:coincidence_erg_dep}. The x and y axes represent energy deposition in E and $ \rm \Delta E$ detectors, respectively, the subtitles of panels indicate the channel combinations.  Having two channels for each of the detectors (see figure~\ref{fig:electronic_sketch}) there are four possible combinations of true coincidence energy deposition in two channels AC, AD, BC and BD. For example, if there is a pulse recorded in channel C then we create entries in both the spectra for channel combinations of A + C and B + C, as shown in panels on the left column of figure~\ref{fig:coincidence_erg_dep}. Same logic was applied to the channels E,F,G and H.  

Three panels corresponding to channel combinations AC, BC and BD contain a prominent energy deposition peak around 10 to 12~MeV (in the E detector) with a "banana"-shaped tail spanning to the left side. The peak corresponds to the energy deposition of protons of energy close to  14~MeV impinging on the detectors and creating a coincidence event. Most of the events in that energy region are induced by protons knocked out from the converter foil due to source neutron interactions and are referred to in the text as the "DT peak". The "banana" tail on the left side of the peak also corresponds to a proton coincidence energy deposition; however, the origin of such events is due to scattered neutron interacting in the converter foil, neutron-induced n,p reactions in the surrounding materials, and random coincidence events. The area around the DT peak together with the "banana" tail corresponds to our region-of-interest  as it contains the proton energy deposition which can be related to the neutron energy. 

The top right panel in figure  \ref{fig:coincidence_erg_dep} corresponding to energy deposition in the detector channels AD which lacks the prominent DT peak feature. For total energy deposition $ \rm \Delta E + E $  above 10~MeV the number of events is at least a factor of 100 smaller compared to other channel combinations. The probability of a proton that originates in the converter foil and creates signal in both channels AD is very small. Figure~\ref{fig:tpr}B indicates that the impinging proton interacting in channel A would need to significantly change direction in the $\Delta$E detector to induce a signal in channel D. The energy deposition in this channel combination is mainly dominated by background events such as neutron-induced nuclear reactions in the detectors, charged particles originating from the surrounding materials, and random coincidences. This observation increases confidence that the background influence on the DT peak is small. 
There is a possibility of energy deposition in three different detector segments, which correspond to different channels, simultaneously. For instance, a single proton can induce a signal in channel combinations such as ABC, BCD, EFG, FGH. Events that induce a signal in channels AB or CD within a time window of 1000~ns and create a coincident signal in the third channel were considered true triple coincidence events. Same logic was applied to the events in channels E,F,G and H.  The energy depositions in channels A, B, or E, F were summed to represent the energy deposited in the $\rm \Delta E$ detector, while energy depositions in channels C, D, or G, H were summed to represent energy deposition in the E detector.

Events that did not satisfy the triple coincidence condition but still induced a signal in three channels were double counted. For example, if a signal is induced in channels ABC but does not meet the conditions of true triple coincidence, it is counted as two separate events: AB and AC. The same logic applies to other channel combinations. This practice may lead to double counting of the same signal and/or misinterpretation of energy deposition. Only 0.04\% events were double counted within the defined region of interest. Consequently, such events are expected to have a negligible influence on the final results.

The energy deposition spectra  (see marks in ~figure~\ref{fig:coincidence_erg_dep}  top left panel) contain artefacts due to the acquisition trigger threshold. 
In addition due to software threshold there are no events below energies 830~keV  and 200~keV for detectors E and $\rm \Delta$ E respectively.

\subsubsection{Anti-Coincidence Measurement}
\label{subsub:anti-coincidence}

\begin{figure}[!ht]
    \centering
    \includegraphics[width=0.95\textwidth]{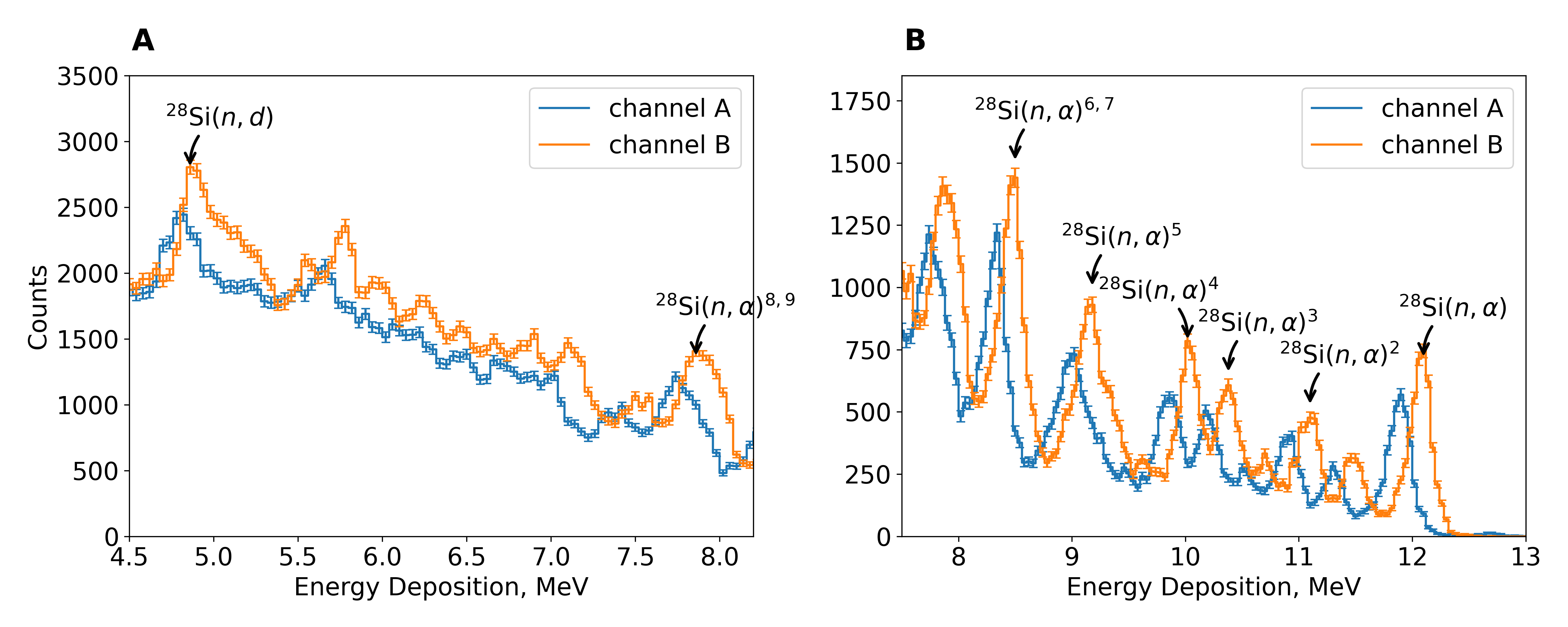}
    \caption{ Example of $\rm ^{28}Si (n,d) $  and $\rm ^{28}Si (n,\alpha) $  reaction signatures in the measured energy deposition in the channels A, B. The energy spectra is constructed assuming no signal is present in other channels simultaneously. The indices i next to $ \rm ^{28}Si (n,\alpha) ^i $ mark the excited state number. x-axis corresponds to energy deposition in $\rm \Delta$E detector and y-axis - to normalised intensity per source particle. Energy bin width is 80~keV.  }
    \label{fig:exp_na_nd}
\end{figure}

Due to small distance of 40~cm between the neutron source and the detectors some of  the source neutrons will penetrate the shielding and induce nuclear reactions in the detectors. This has been observed and an example of anti-coincidence energy deposition spectra for the $\rm \Delta E$ detector channels is shown in figure~\ref{fig:exp_na_nd}. For example, channel A in figure~\ref{fig:exp_na_nd}A corresponds to events in channel A when there is no signal in channels B,C and D.  

All channels in anti-coincidence spectra have a peak in the energy range of  4.75 - 5.00~MeV and  11 - 12~MeV. The most likely candidates are nuclear reactions n,d and $ \rm n,\alpha $ in silicon with reaction thresholds \textit{Q} of 9.36~MeV and 2.65~MeV, respectively.  Assuming all of the products deposit their energy within the same active detector volume and a peak neutron energy of 14.08~MeV we estimate that reaction $\rm ^{28}Si (n,d) $  and $\rm ^{28}Si (n,\alpha) $ should deposit energy of 4.72~MeV and 11.43~MeV which is within the range of our observations. The peaks in figure~\ref{fig:exp_na_nd}B in energy range 8 - 11~MeV are likely originating as well from $\rm ^{28}Si (n,\alpha) $  reaction when compound nuclei decays into an excited state nuclide, consequently the energy deposition peaks shift to lower energies. Analogous observations have been published previously~\cite{REBAI2019162637,ruddy_fast_2006} and correspond well with current o. The observed deviation from the expected energy deposition could be explained by uncertainty in the energy resolution calibration (shaded band in the figure~\ref{fig:erg_calib}B). We observe that deviation is larger for higher energy deposition, which is reasonable as 11.43~MeV it is outside the range of performed energy calibration 3-7~MeV.  The results for channels C and D are very similar, however due to poorer energy resolution the $\rm ^{28}Si (n,\alpha) ^{2,3,4,8,9}$ reaction peaks were not fully resolved. 


\newpage
\section{Geant4 simulations of the experiment}
\label{sec:simulations}

A Geant4 model of the experiment was setup based on the previous research~\cite{Marcinkevicius2019,Marcinkevicius2018}. The purpose of the model was to investigate whether it could replicate the experimental results. Additionally we have used the model to determine the region-of-interest (ROI) for the proton energy deposition.  Discriminating proton energy deposition from other particle influences, using a ROI, could possibly improve the signal-to-background ratio~\cite{Marcinkevicius2019}. 
The sections below provide a description of the Geant4 model implementation, in particular the neutron source description and the experiment geometry. The authors also reviewed the simulation results on coincidence spectra, region-of-interest quantification, and anti-coincidence spectra. 

\subsection{Model Changes}
  Due to the proximity of the spectrometer to the neutron source, it was necessary to model the distribution of the source in 4$ \pi$ instead of using a collimated neutron beam, as has been done previously~\cite{Marcinkevicius2019,Marcinkevicius2018}. This has significantly increased the simulation time and it was decided to stop the neutron transport calculations  for neutron energies <50~keV which has significantly reduced computational complexity. This energy cut-off should have little influence on the comparison with experimental results due to a  experimental trigger threshold being above 50 keV. 
 
\subsection{Neutron Source}

 It was assumed that the DT neutron source emits neutrons uniformly in all directions. The neutron energy spectrum was calculated for a deuterium beam of 120~keV energy impinging on a Tritium enriched target and emitting neutrons at 90$\rm ^o$ to the beam direction. This assumption is considered reasonable as the spectrometer line-of-sight is perpendicular to the beam direction and the converter foil would cover angles from 86.8$\rm  ^o$  - 93.2$\rm  ^o$. The number of source particles launched in the simulation was $ 2 \times 10^{11}$. 

\begin{figure}[!ht]\centering
	\includegraphics[width=0.95\textwidth]{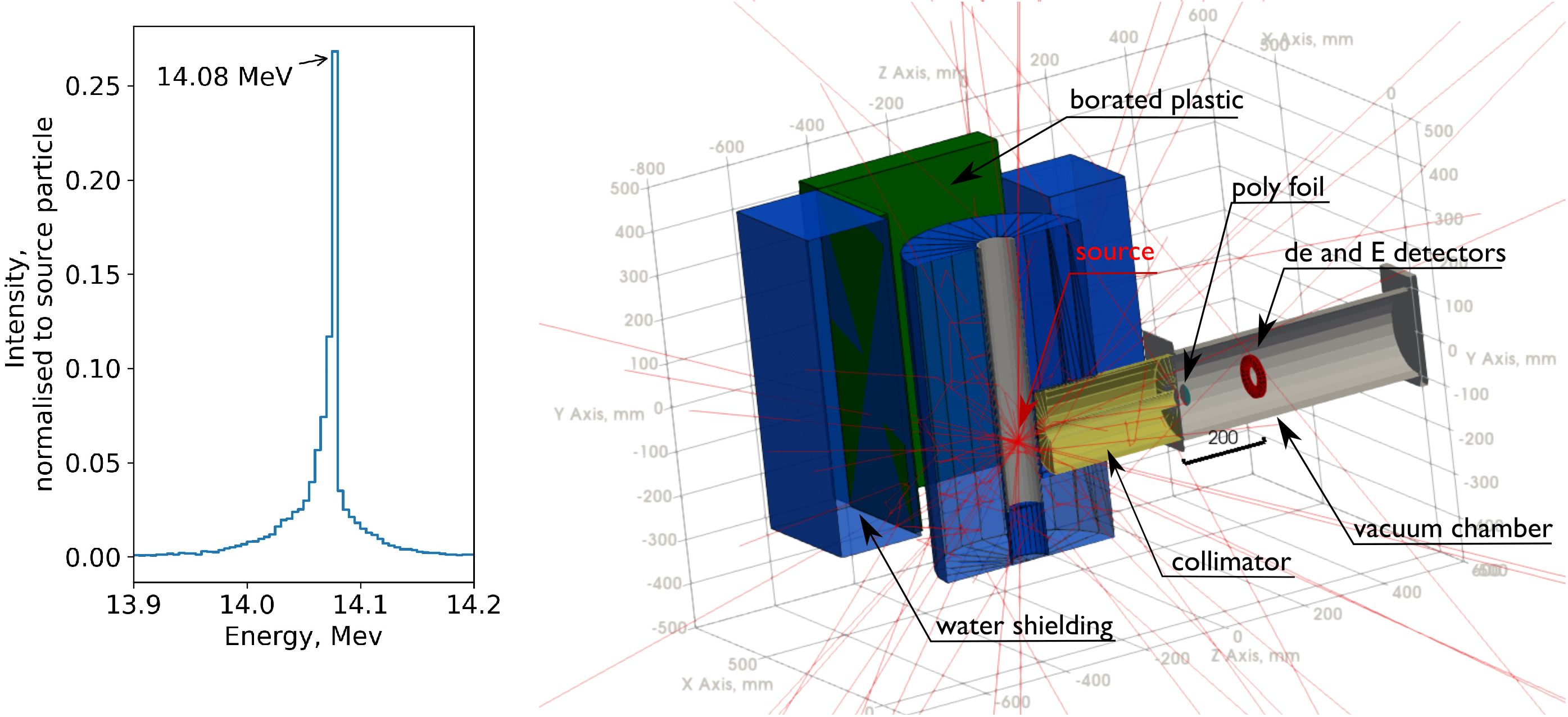}
	\caption{ Left panel: neutron source energy distribution. Right panel: A  representation of the Geant4 model geometry with scale; the geometry is sliced at plane x=0; the red spot with red lines represents neutron tracks starting in the source position.}
	\label{fig:g4_geo}
\end{figure}

\subsection{Geometry}
The geometry of the Geant4 model is shown in figure~\ref{fig:g4_geo} right panel. The location of the source is marked by a spot  (red) on the left-hand side of the geometry. The source is placed in a cylindrical hole inside a container filled with water (dark blue). The paraffin collimator (shade red) is extending up to the vacuum chamber of the spectrometer (grey). The spectrometer converter-foil (cayen) is followed by two silicon detectors (deep red). In addition, the foil is surrounded by a thin ring of steel that represents a foil holder. The vacuum vessel has been implemented as an aluminum tube with steel (left) and plastic (right) lids. The vacuum chamber was filled with low-density air (1.2e-9 $ \rm g/cm^{-3}$ ) to account for the attenuation of the charged particles before reaching the detectors.

\subsection{Results}
\label{sub:G4_results}

The intrinsic resolution of the detector was taken into account in the simulations by broadening the energy deposition according to the measured energy resolution. The energy deposition in the  $ \Delta$E and E detectors were folded with Gaussian function of  85~keV  and 255~keV  standard deviation respectively, independent of energy deposition. Cross-correlations between the segments in the same detector were not taken into account. For example, if particle at the detector surface crosses into detector $\rm \Delta E$ and induces signal in both channels A and B  all energy deposition would be attributed to channel A.

\subsubsection{Coincidence Spectra}
\begin{figure}[!ht]\centering
	\includegraphics[width=0.99\textwidth]{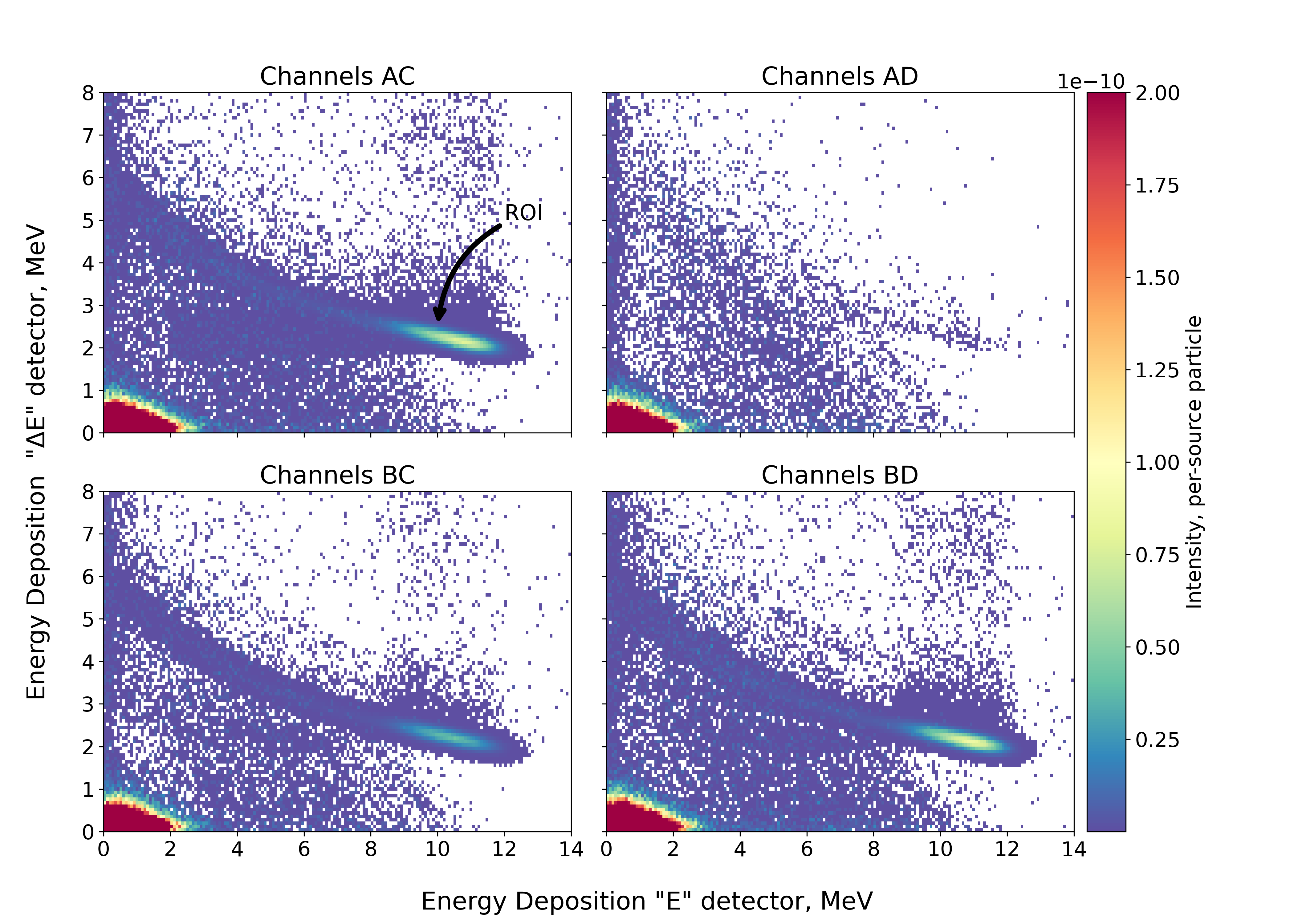}
	\caption{Simulation results of energy deposition in the spectrometer for different channel coincidences. The results are normalised to the total source particles simulated. ROI - marks region of interest where
proton energy deposition dominates.  Energy bin width is 100~keV.} 
	\label{fig:G4_2dhist}
\end{figure}

The modelled energy coincidence spectra are presented in figure~\ref{fig:G4_2dhist}, where  events contributing energy to both of the detectors are depicted. The panels in the figure~\ref{fig:G4_2dhist}  correspond to the previously  discussed figure~\ref{fig:coincidence_erg_dep}, where the panel titles indicate the channels of coincidence.  The features of the spectra are similar to the experimental observations; the "DT peak" is one of the dominant features in channel combinations of AC, BC and BD. There is an absence of events in ROI in coincidence channels AD (figure~\ref{fig:G4_2dhist} top right panel). There is a strong contribution to the energy spectra spanning from 0 - 2~MeV in the E detector. The origin of this characteristic was not investigated as it was far from ROI and was not observed in the experiment; however, gamma-induced events would be a significant contributor in the energy range below 2~MeV.

\newpage
\subsubsection{ROI selection}
The simulation results were used to estimate the ROI where energy deposition is dominated by protons. The ability to separate the  proton-induced signal from other contributions would significantly improve the signal-to-background ratio under experimental conditions The ROI was estimated using the same principles as in ~\cite{Marcinkevicius_2019}.

\begin{figure}[!ht]\centering
	\includegraphics[width=0.99\textwidth]{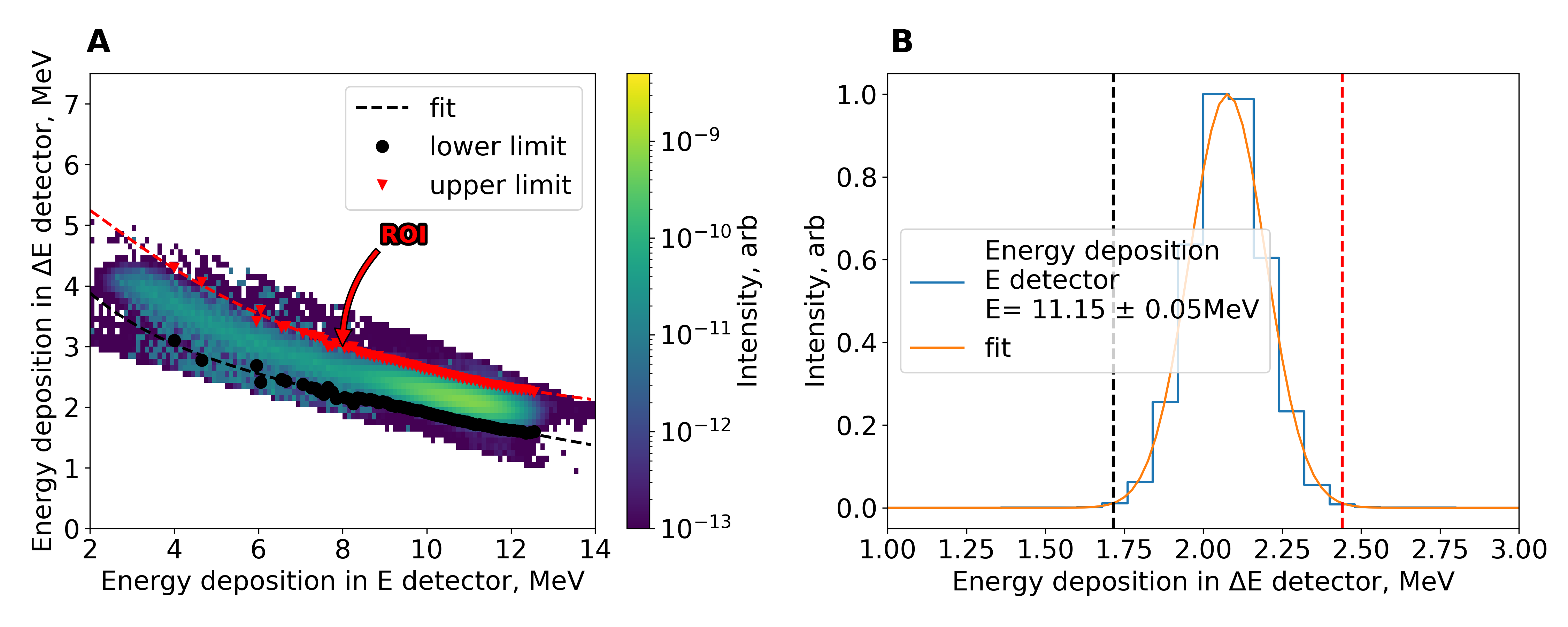}
	\caption{Geant4 simulation results of energy deposition in the $\Delta$E-E detectors and estimation of the ROI. A - an energy deposition in  $\Delta$E-E detectors as simulated with Geant4 with applied constrains for the fitting procedure, bin width of 100~keV. Each marker labelled as "lower limit" or "upper limit" correspond to the Gaussian fit results.  The dashed lines mark an exponential approximation for the ROI bounds.  Note the logarithmic scale on the intensity axis. Panel  B shows an example of  bounds estimation for one energy bin  (11.15 $\pm$ 0.05 MeV) in E detector. The dashed lines (3 $\sigma$)  mark upper and lower bounds and the colour coding correspond to the left panel upper and lower limit respectively. 
	 Each fit provides one point for the ROI upper and lower limits.}
	\label{fig:G4_banana}
\end{figure} 
The Figure~\ref{fig:G4_banana}~A shows the coincidence energy deposition in both detectors, upper and lower bounds estimated from the energy deposition in the $ \rm \Delta E$ detector and the parametrisation of the ROI. 
Figure~\ref{fig:G4_banana}~B  shows an example of how the upper and lower bounds were estimated for one slice in the E detector (here E = 11.15$\pm $0.05). A Gaussian  function was fit to the  energy deposition in the $\Delta$E detector, the $ \pm$3$\sigma$ of the fit were set as an upper and lower bounds of the ROI as shown in example in figure~\ref{fig:G4_banana}~B . Such a procedure was repeated for all of the detector E slices up to energy deposition of 12.8~MeV in order to estimate the bounds in the energy region of interest and thus set the ROI; only fits with reduced chi-square > 0.9 were used. The energy limit of 12.8~MeV corresponds to the maximum intensity of proton energy deposition.  The calculated bounds were then parameterised according to equation~\ref{eq:exponential_approximation}, where A,B,C,D,E,F,G are constants,  $\Delta E$ corresponds to the energy deposition in detector $\rm \Delta E$ and E to the energy deposition in detector E. 
\begin{equation}
\label{eq:exponential_approximation}
    E = A \times e^{(B - C \times \Delta E )} + D + E \times e^{(F - G \times \Delta E )}
\end{equation}
A time coincidence between 0.18 to  0.3~ns  and  energy constraints were applied,  as depicted in the Figure~\ref{fig:G4_banana},   during the analysis to select the proton-induced signal from the simulations and to limit the background influence on the Gaussian fit.  Only one ROI was constructed for the whole spectrometer, however there is a possibility to create ROI for each possible channel combination.

\subsubsection{Anti-coincidence spectra}
\begin{figure}[!ht]\centering
    \centering
    \includegraphics[width=0.99\textwidth]{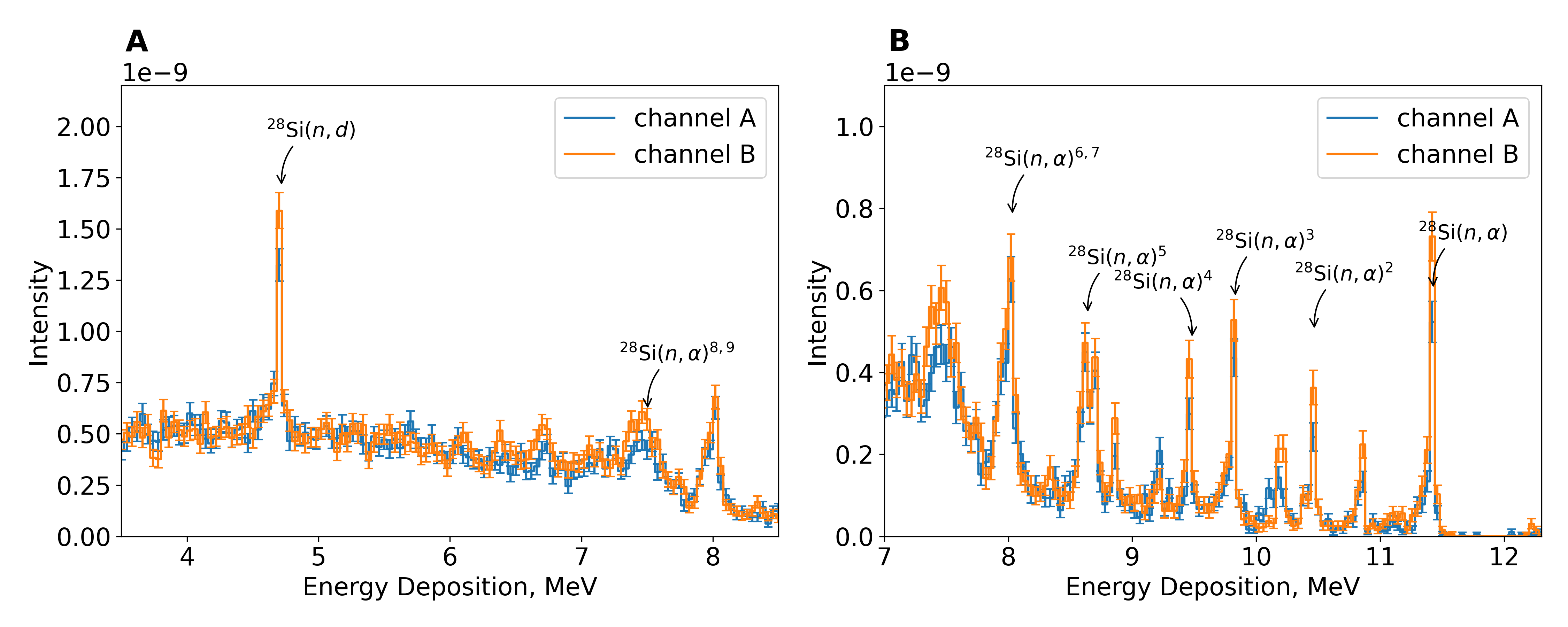}
    \caption{ Example of $\rm ^{28}Si (n,d) $  and $\rm ^{28}Si (n,\alpha) $  reaction signatures in the simulated energy deposition in the channels A, B. The energy spectra is constructed assuming no signal is present in other channels simultaneously (anti-coincidence spectra). x-axis corresponds to energy deposition in $\Delta$E detector and y-axis - to normalised intensity per source particle. Energy bin width 50~keV. Error bars represent statistical error only.  }
    \label{fig:g4_na_nd}
\end{figure}

Similarly to the experiment we have observed both n,d and n, $ \alpha $ reaction signatures in the energy deposition spectra in the simulation results (see figure~\ref{fig:g4_na_nd}).  To investigate this, anti-coincidence energy spectra were constructed without adding broadening due to detector energy resolution to determine accurate peak positions. The results suggest that the peak positions are the same in all detector channels. The energy deposition corresponds well to the expected values of 4.72~MeV and~
 11.43~MeV as discussed in section~\ref{subsub:anti-coincidence}. Furthermore, according to the particle tracks recorded, the energy deposition in these energy ranges was dominated by deuterons and $\alpha$ particles, respectively. 
This confirms the previously stated claim that the peak origin is neutron-induced reactions (n, d) and $ \rm (n,\alpha)$ in $ \rm ^{28}Si $.

\subsection{Energy Calibration Correction}
\begin{figure}[!ht]\centering
    \centering
    \includegraphics[width=0.99\textwidth]{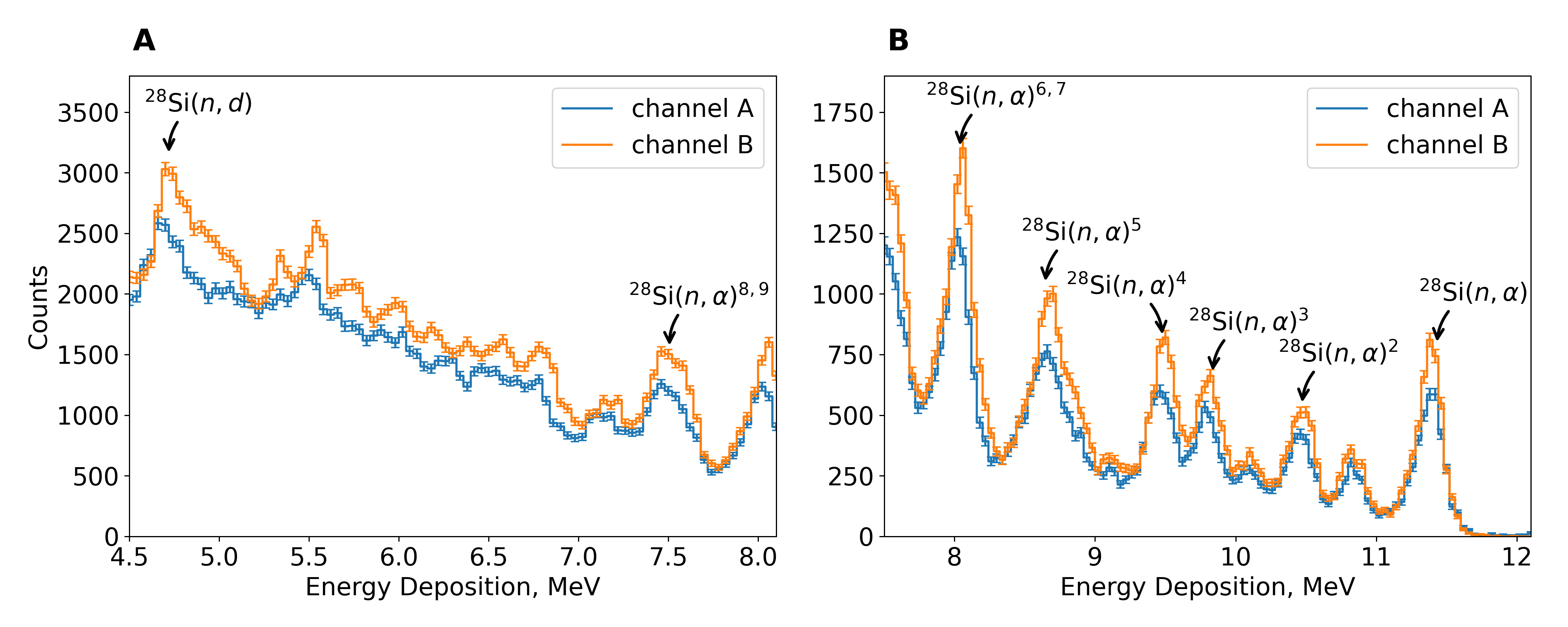}
    \caption{ Example of the experimental anti-coincidence energy deposition in the $\rm \Delta E$ detector after energy calibration correction. The energy bin width 50~keV. Error bars represent statistical error only.}
    \label{fig:corrected_spectra}
\end{figure}

The observation of nuclear reaction signatures was used to correct the energy calibration. The energy deposition was shifted according to a new calibration function for each of the detector channels, where the experimentally identified reaction peaks were shifted to the values determined in the
simulations, i.e., 4.72 MeV and 11.43 MeV.  The new calibration function is defined in equation~\ref{eq:erg_correction_linear_fit} , where A is a constant, k is the slope, $E_{true}$ is the energy deposition derived from the simulations of nuclear reactions (figure~\ref{fig:g4_na_nd}), $E_{dep}$ is a measured energy deposition (figure~\ref{fig:exp_na_nd}) and index i corresponds to the detector channel. 
\begin{equation}
     E_{true, i} = A_i + k_i \times E_{dep, i}
     \label{eq:erg_correction_linear_fit}
\end{equation}
The resulting corrected experimental anti-coincidence energy spectra are shown in figure~\ref{fig:corrected_spectra} where we can observe a much better peak alignment well within experimental detector resolution. 
\newpage

\section{ Comparison between Model and Experimental results}
\label{sec:results}

\begin{figure}[!ht]\centering
	\includegraphics[width=0.48\textwidth]{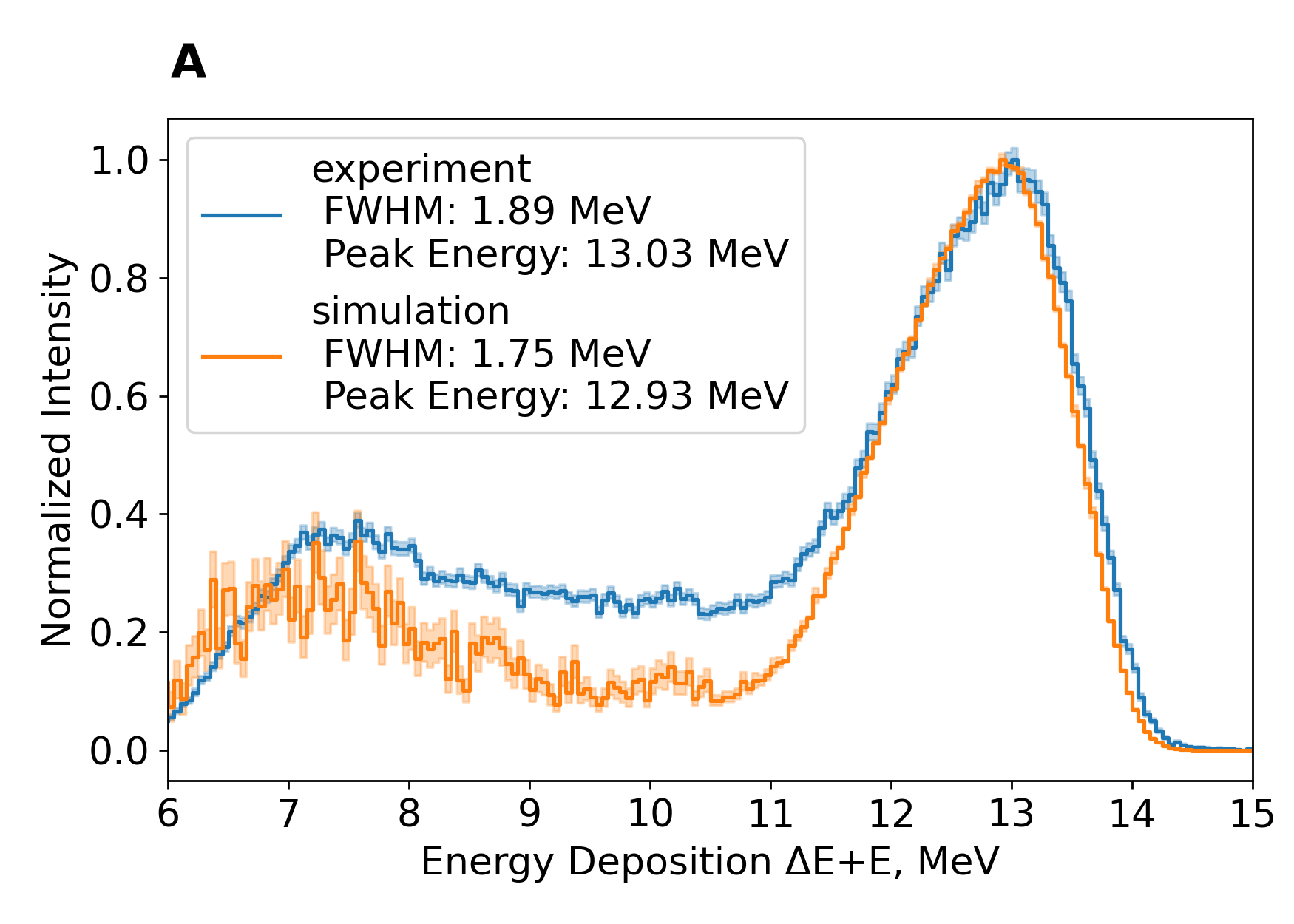}
 \includegraphics[width=0.48\textwidth]{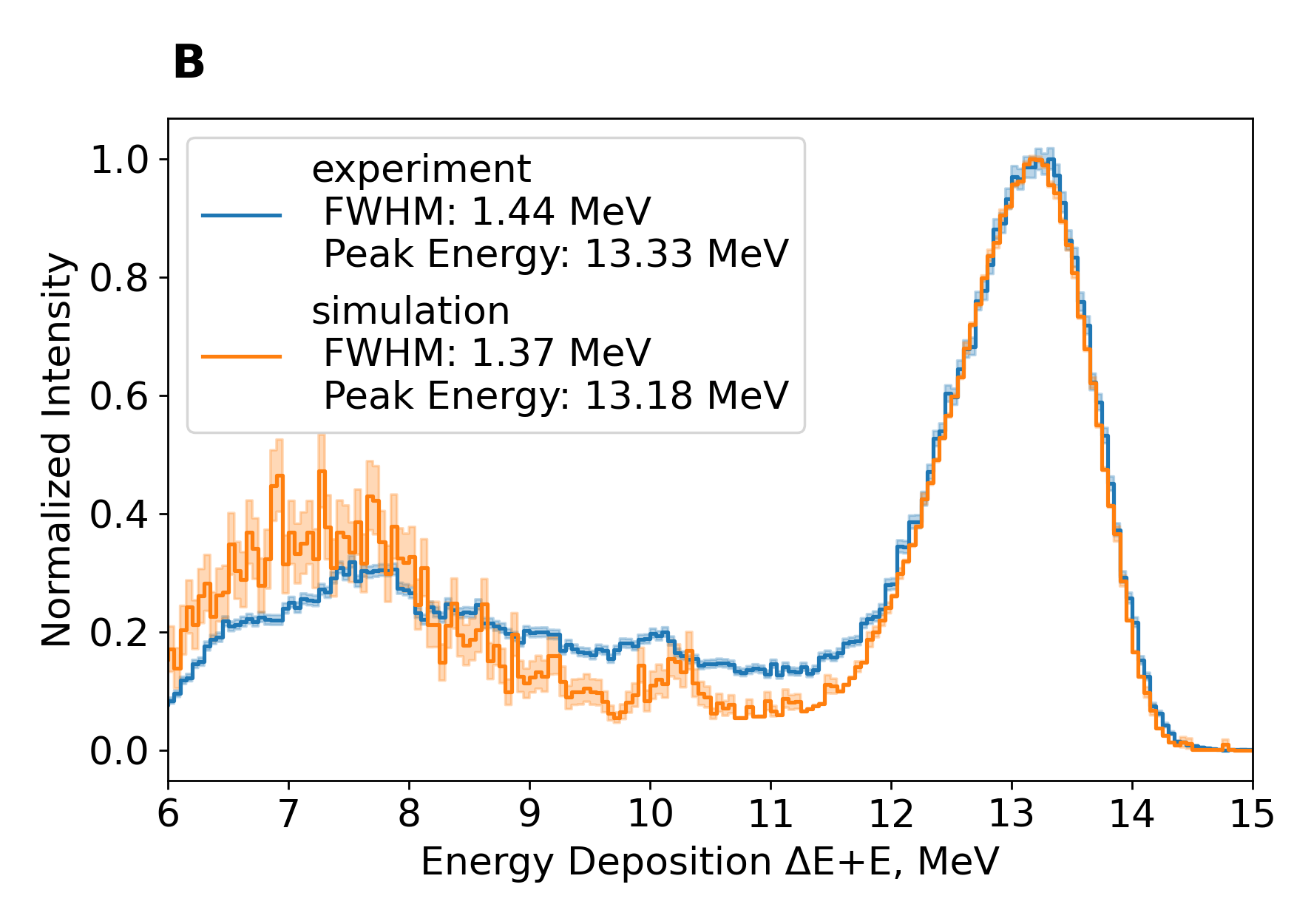}
	\caption{Total energy deposition ($ \rm \Delta E + E$) of the coincidence events (within 7500~ns) in the Region of interest. Comparison of experimental (solid blue line histogram) and simulation results. Both experimental and simulation data are normalised to unity at their respective  maximum for better shape comparison. Energy bin width 50~keV.  Shaded are represents statistical error only.}
	\label{fig:exp_vs_model}
\end{figure}

The total energy deposition spectra in both $ \rm \Delta E$ and E  detectors is compared in the 
 Figure~\ref{fig:exp_vs_model}, only the events within the previously defined ROI are shown. The time coincidence window used is 7500~ns. The energy spectra are normalised to the maximum for a better shape comparison.
 The determined maximum of the peak and FWHM are noted next to the experiment and simulation labels. The panel A shows results collected with the digitzer 3920, panel B results collected with the digitizer 0901.  The simulation results have passed  convergence tests employed \cite{kordusas_population_convergence_tests}.
 
  The shapes of the energy deposition spectra are very similar, especially in the energy range above 11~MeV. The mean energy deposition in the peak  as well as full width half maximum (FWHM) match within the measured detector energy resolution.  The spectra shape matches very well at the high energy tail, which is crucial in fusion diagnostics to determine different fusion component intensities \cite{Carl2015}. However, the simulation results systematically underestimate the experimental data in terms of both energy of the peak at the maximum intensity as well as FWHM. The tail on the left hand side corresponds to background events and has bigger mismatch in shape, and in part explains the differences of FWMH between the simulations and the experiment.  The simulations were not designed to replicate the full background contribution in the lower energy range, consequently a more detailed model of the surroundings would be necessary to have a better match. 
 
 The ratio between the simulated and measured spectrometer efficiency (C / E) is 0.78 and 0.65 for results presented in panels A and B. The efficiency was estimated as an integral of the peak above 11~MeV where the low energy tail contribution is limited, the statistical uncertainty for both experiment and simulation results are below 1\%.  The observed differences between the experiment and simulation are likely due to the uncertainties on distance between the source and the spectrometer and collimator alignment. The difference C/E ratio for the two digitizers also indicates issues with  alignment, otherwise we would expect the  C/E be contstant for both of the digitizers. Furthermore, no calibrated source yield monitor was present during the experiment. In general, an accurate calibration of the spectrometer efficiency would require an improved control of experimental conditions.

\begin{figure}[!ht]
    \centering
    \includegraphics[width=0.99\textwidth]{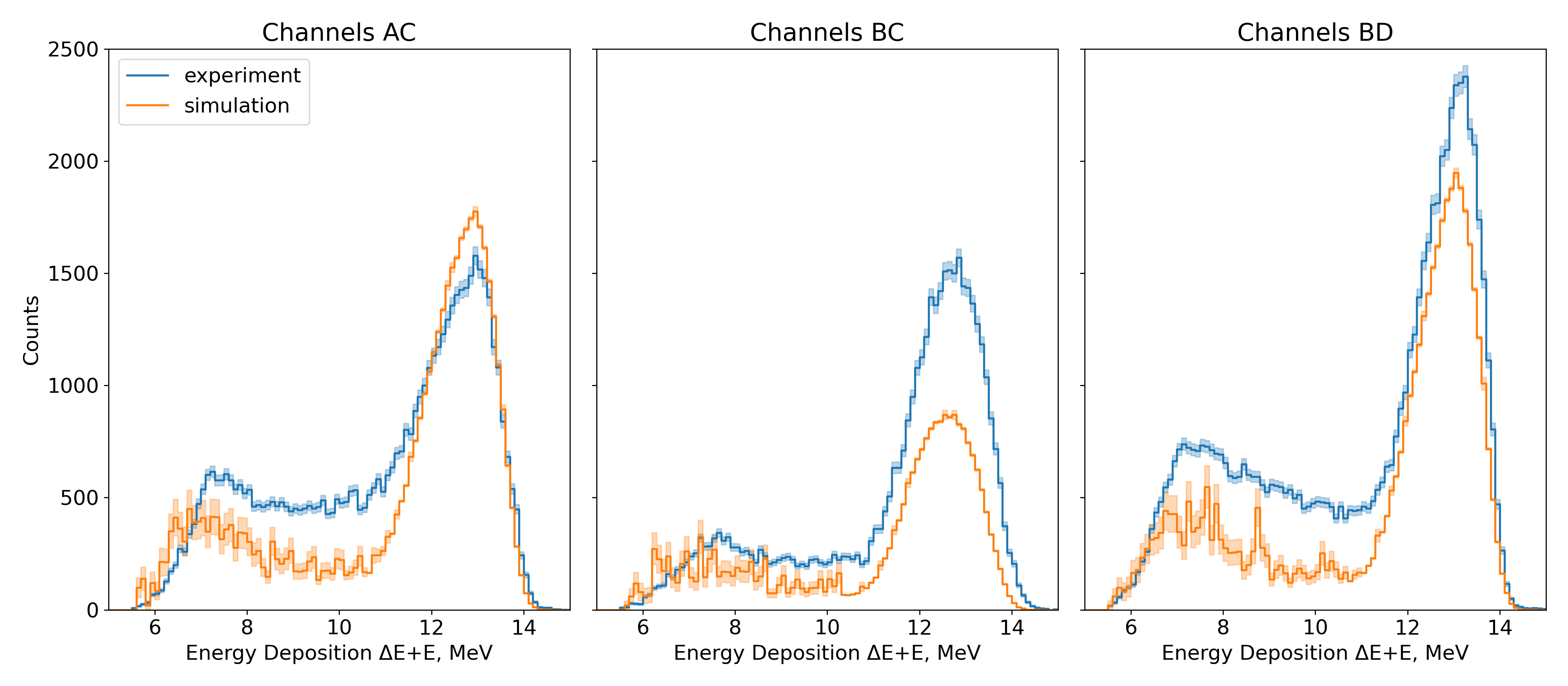}
    \includegraphics[width=0.99\textwidth]{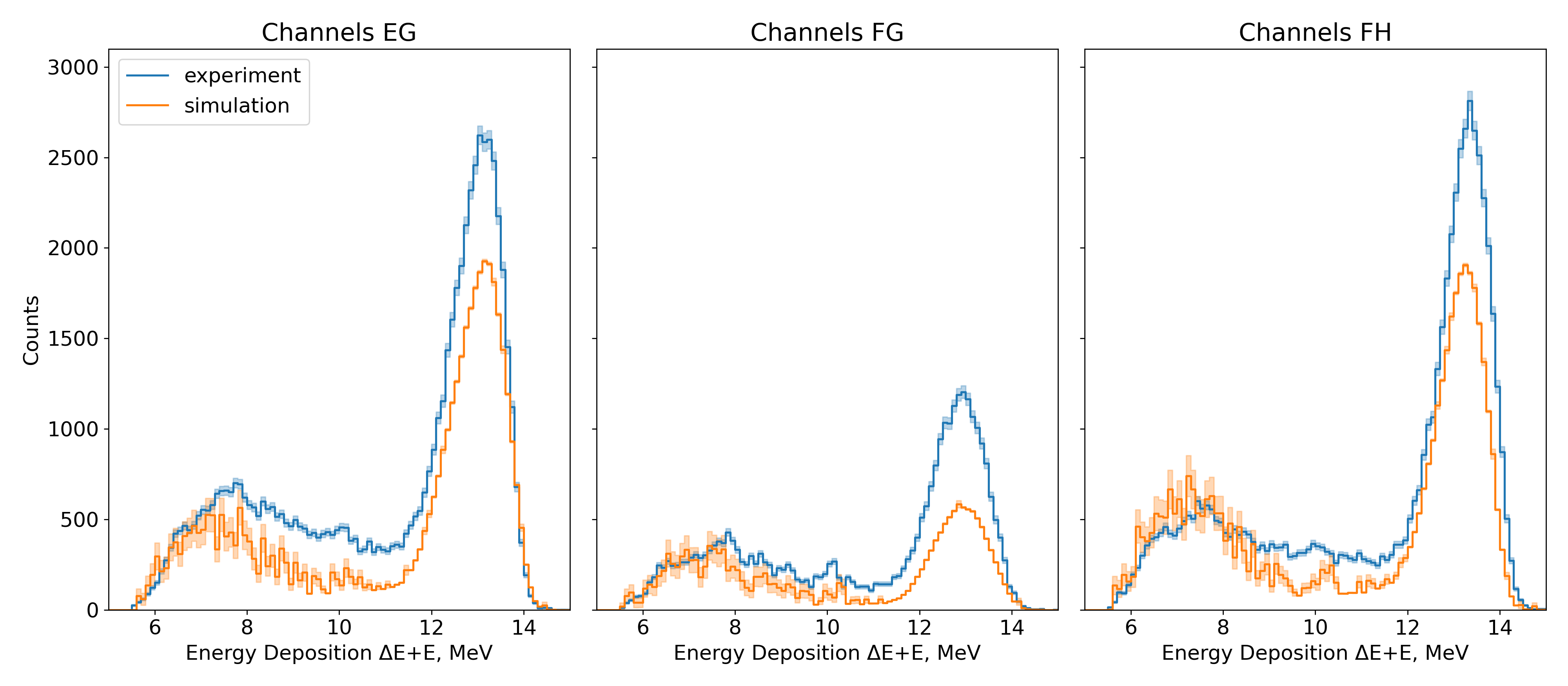}
    \caption{Total energy deposition ($ \rm \Delta E + E$) of the coincidence events (within 7500~ns) in the ROI for channel combinations of AC, BC, BD, EG, FG and FH as indicated by subtitles. Comparison between experimental and simulation results. Simulation results are normalised to the estimated total source strength and acquisition time. Energy bin width 100~keV. Shaded area represent statistical error only. }
    \label{fig:exp_vs_model_by_channels}
\end{figure}

Furthermore, Figure~\ref{fig:exp_vs_model_by_channels} shows the differences between simulations and experiment for each channel combination. 
The figure compares the total number of counts per energy bin measured by the experiment with simulation results normalised to the estimated total source strength and acquisition time. There is an indication that the efficiency mismatch is strongly dominated in the channel combinations of BC, FG and further investigation is necessary to quantify  the source of this deviation; a possible reason could be inaccuracies in the distance between the detectors or alignment with the source. Overall we can conclude that our model can replicate the experimental result shape rather well for the whole detector (as shown in figure \ref{fig:exp_vs_model}), however there are some discrepancies between simulated and experimental spectrometer efficiency and separate channel contributions.

\section{Conclusions and Discussion}
\label{sec:conclude}

We have measured energy deposition spectra with a prototype thin-foil proton recoil $ \rm \Delta E-E $ spectrometer using a DT neutron source, and we have performed calculations based on a previously developed Geant4 model~\cite{Marcinkevicius2019,Marcinkevicius2018}. 
In general, the measurement and simulation results shape match well within the energy range of interest (above 11~MeV).  The measured energy deposition peak in the spectra associated with neutrons form a DT source is replicated well by our model, the FWHM difference are below 120 keV. The simulated spectrometer detection efficiency is underestimated compared experimental results up to 35\%, depending on the selected channels, for an energy range above 11~MeV.  We have also demonstrated that background reduction methods~\cite{Marcinkevicius_2019}  based on region-of-interest can be successfully used to reduce the background contribution in the final energy spectra. 
A better match between the experiment and simulations would require better controlled experimental conditions. The mismatch in the low energy tail of the energy spectra  is likely due to model simplifications.

Furthermore, we have observed neutron induced reactions  $\rm ^{28}Si (n,d) $  and $\rm ^{28}Si (n,\alpha) $ in both experiment and simulations, similar to observations reported for silicon carbide detectors~\cite{REBAI2019162637,ruddy_fast_2006}. The high energy threshold of these reactions and peak separation from the background enabled us to successfully improve the energy calibration of the Si detectors used in the experiment. 

In addition, the reactions involving $\rm ^{28}Si$ could be used to monitor the changes of the detector energy resolution and energy calibration during the experimental campaign.  This is relevant for experiments like ITER when access to the spectrometer could be limited for  long periods of time. 
The application of this method at ITER may be limited due to multiple differences from the current experimental set-up. At ITER the spectrometer would be much better shielded from direct neutron irradiation, due to a longer collimator and more
substantial radiation shielding~\cite{scholz_conceptual_2019}; thus most nuclear reactions in the silicon detectors would be induced by neutrons scattering in the conversion foil itself. Consequently, the peak energy, shape, and intensity may change. The advantages at ITER would be a much higher detection count rate, smaller background contribution and significantly larger data set. A thorough investigation should be performed to quantify possibilities of such method at ITER.

The spectrometer experiment has yielded two main outcomes: firstly, it confirmed the practicality of employing an annular $ \rm \Delta E-E $ spectrometer system with segmented silicon detectors for neutron diagnostics in DT reaction-based sources, and secondly, it provided an empirical validation for previously developed Geant4 simulations model~\cite{Marcinkevicius2019,Marcinkevicius2018} of the spectrometer. These findings collectively highlight the spectrometer applicability in future fusion devices like ITER for neutron diagnostic purposes. Furthermore, our observations of $\rm ^{28}Si (n,d) $  and $\rm ^{28}Si (n,\alpha) $  reactions show a potential  for monitoring of silicon detector efficiency and energy calibration during the experiment.

Future investigations could expand on this work. Simulations of further prototype tests could be improved by a more complete account of the influence of scattered neutrons. Additionally, an experiment parameter study could be performed to propagate the experiment set-up uncertainties to the simulated energy deposition. New hardware features could be added to the experimental set-up; for example, a shaping amplifier and dedicated timing circuit could significantly improve the precision in the coincidence measurement. The spectrometer could be scaled up according to previous studies~\cite{scholz_conceptual_2019, Marcinkevicius_2019} to be applied for diagnostic purposes in high yield DT fusion devices such as ITER. Additionally future experiments should rely on independent neutron rate monitors for better absolute normalisation. 
\section{Acknowledgements}

This work has been carried out within the framework of the EUROfusion Consortium and has received funding from the Euratom research and training programme 2014-2018 and 2019-2020 under grant agreement No 633053. The views and opinions expressed herein do not necessarily reflect those of the European Commission.

I would like to extend my gratitude towards the Lund University Nuclear Physics Division and especially Mikael Elfman for providing the opportunity and facilities to perform the experiment using the neutron source. 

\bibliographystyle{unsrt}  
\bibliography{References.bib}

\end{document}